\documentclass[aps, prd,twocolumn,showpacs]{revtex4}
\usepackage{graphicx}
\usepackage{latexsym}
\usepackage{amsmath}
\usepackage{psfrag}
\usepackage{amssymb}
\usepackage{color}

\newcommand{\bq}{\begin{equation}}
\newcommand{\eq}{\end{equation}}
\newcommand{\bqn}{\begin{eqnarray}}
\newcommand{\eqn}{\end{eqnarray}}

\begin{document}
\title{Interacting holographic tachyon model of dark energy}
\author{Alberto Rozas-Fern{\'a}ndez$^1$}
\email{a.rozas@cfmac.csic.es} \affiliation{Colina de los Chopos,
Instituto de F{\'i}sica Fundamental, Consejo Superior de
Investigaciones Cient{\'i}ficas, Serrano 121, 28006 Madrid, Spain}
\author{David Brizuela$^2$}
\email{d.brizuela@uni-jena.de}
\affiliation{Theoretisch-Physikalisches Institut, Friedrich-Schiller-Universit\"at,
Max-Wien-Platz 1, 07743 Jena, Germany}
\author{Norman Cruz$^3$}
\email{norman.cruz@usach.cl} \affiliation{Departamento de
F{\'i}sica, Facultad de Ciencia, Universidad de Santiago de Chile,
Casilla 307, Santiago, Chile}

\date{\today}

\begin{abstract}
We propose a holographic tachyon model of dark energy with
interaction between the components of the dark sector. The
correspondence between the tachyon field and the holographic dark
energy densities allows the reconstruction of the potential and
the dynamics of the tachyon scalar field in a flat
Friedmann-Robertson-Walker universe. We show that this model can
describe the observed accelerated expansion of our universe with a
parameter space given by the most recent observational results.
\end{abstract}

\maketitle

\section{Introduction}

Recent cosmological observations from Type Ia supernovae (SN Ia)
\cite{SN}, Cosmic Microwave Background (CMB) anisotropies measured
with the WMAP satellite \cite{CMB}, Large Scale Structure
\cite{LSS}, weak lensing \cite{WL} and the integrated Sach-Wolfe
effect \cite{ISWE} provide an impressive evidence in favor of a
present accelerating Universe. Within the framework of the
standard Friedmann-Robertson-Walker (FRW) cosmology, this present
acceleration requires the existence of a negative pressure fluid,
dubbed dark energy (DE), whose pressure $p_{\Lambda}$ and density
$\rho_{\Lambda}$ satisfy
$\omega_{\Lambda}=p_{\Lambda}/\rho_{\Lambda}<-1/3$. In spite of
this mounting observational evidence, the underlying physical
mechanism behind this phenomenon remains unknown. Interesting
proposals are the quantum cosmic model \cite{benigner} and $f(R)$
theories (see \cite{f(R)} for recent reviews and references
therein). Likewise, we have a plethora of dynamical dark energy
models \cite{dynamical dark energy}.

On the other hand, based on the validity of effective local
quantum field theory in a box of size $L$, Cohen et al
\cite{Cohen:1998zx} suggested a relationship between the
ultraviolet (UV) and the infrared (IR) cutoffs  due to the limit
set by the formation of a black hole. The $UV-IR$ relationship
gives an upper bound on the zero point energy density,
\begin{equation}\label{eq1}
\rho_{\Lambda}\leq L^{-2}M_{p}^{2},
\end{equation}
where $L$ acts as an IR cutoff and $M_p$ is the reduced Planck
mass in natural units. This means that the maximum entropy in a
box of volume $L^{3}$ is
\begin{equation}\label{eq2}
S_{max}\approx S^{3/4}_{BH},
\end{equation}
being $S_{BH}$ the entropy of a black hole of radius $L$. The
largest $L$ is chosen by saturating the bound in Eq. (\ref{eq1}) so
that we obtain the holographic dark energy density
\begin{equation}
\rho_{\Lambda}= 3c^{2}M^{2}_{p}L^{-2},
\end{equation} where c is a free dimensionless ${\cal O}(1)$ parameter
and the numeric coefficient is chosen for convenience.
Interestingly, this $\rho_{\Lambda}$ is comparable to the observed
dark energy density $\sim10^{-10}eV^{4}$ for the Hubble parameter
at the present epoch $H=H_{0}\sim10^{-33}eV$.

If we take $L$ as the Hubble scale $H^{-1}$, then the dark energy
density will be close to the observational result. However, Hsu
\cite{Hsu:2004ri} pointed out that this does not lead to an
accelerated universe. This led Li \cite{Li:2004rb} to propose that
the IR cut-off $L$ should be taken as the size of the future event
horizon of the Universe
\begin{equation}
R_{\rm eh}(a)\equiv a\int\limits_t^\infty{dt'\over
a(t')}=a\int\limits_a^\infty{da'\over Ha'^2}~,\label{eh}
\end{equation}
where $a$ is the scale factor of the universe and $t$ the cosmic
time. Choosing the future event horizon as the $UV$ cut-off
tacitly assumes the acceleration of the expansion of the universe.
Since the accelerating universe is a well supported observational
fact, we believe that this assumption is plausible.

This allows to construct a satisfactory holographic dark energy
(HDE) model which presents a dynamical view of the dark energy
which is consistent with observational data \cite{HDE data}. As a
matter of fact, a time varying dark energy gives a better fit than
a cosmological constant according to some analysis of astronomical
data coming from type Ia supernovae \cite{dynamical DE data}.
However, it must be stressed that almost all dynamical dark energy
models are settled at the phenomenological level and the HDE model
is no exception in this respect. Its advantage, when compared to
other dynamical dark energy models, is that the HDE model
originates from a fundamental principle in quantum gravity, and
therefore possesses some features of an underlying theory of dark
energy.

A further development was to consider a possible interaction
between dark matter (DM) and the HDE \cite{Interacting HDE}.

It is usually assumed that both DM and DE only couple
gravitationally. However, given their unknown nature and that the
underlying symmetry that would set the interaction to zero is
still to be discovered, an entirely independent behavior between
the dark sectors would be very special indeed. Moreover, since DE
gravitates, it must be accreted by massive compact objects such as
black holes and, in a cosmological context, the energy transfer
from DE to DM may be small but non-vanishing. In addition, it was
found that an appropriate interaction between DE and DM can
influence the perturbation dynamics and affect the lowest
multipoles of the CMB angular power spectrum \cite{Interaction in
perturbation dynamics, Interaction in the CMB multipoles}. Thus,
it could be inferred from the expansion history of the Universe,
as manifested in several experimental data. Furthermore it was
suggested that the dynamical equilibrium of collapsed structures
such as clusters would be modified due to the coupling between DE
and DM \cite{Bertolami:2007zm, Kesden}. Most studies on the
interaction between dark sectors rely either on the assumption of
interacting fields from the outset
\cite{Amendola:2000uh,Das:2005yj}, or from phenomenological
requirements \cite{Phenomenological interaction}. The aforesaid
interaction has also been considered from a thermodynamical
perspective \cite{Wang:2007ak, Pavon:2007gt} and has been shown
that the second law of thermodynamics imposes an energy transfer
from DE to DM.

As is well known, the scalar field models are an effective
description of an underlying theory of dark energy. Scalar fields
naturally arise in particle physics including supersymmetric field
theories and string/M theory. However, these fundamental theories
do not predict their potential $V(\phi)$ uniquely. Consequently,
it is meaningful to reconstruct the potential $V(\phi)$ of a dark
energy model possessing some significant features of the quantum
gravity theory, such as the interacting HDE (IHDE) model.

In this  Letter we would like to extend the previous work done by
Zhang et al \cite{Zhang:2007es}, where they took advantage of the
successful HDE model and used the tachyon scalar field as an
effective description of an underlying theory of dark energy, by
incorporating a possible interaction between DM and DE. The
holographic tachyon model of dark energy was also investigated in
\cite{Setare:2007hq} and the interacting tachyon dark energy was
first studied in \cite{Herrera:2004dh}. Tachyonic fields have the
attractive feature that may describe a larger variety of
cosmological evolutions than quintessence fields
\cite{Gorini:2003wa}. Other relevant works on interacting and
non-interacting holographic dark energy can be found in
\cite{Setare:2006wh, Setare:2006sv, Setare:2007jw, Setare:2007mp,
Setare:2007zp}.

The rest of the paper can be outlined as follows. In Sec. II we
build the interacting holographic tachyon model and plot the
potential and the evolution of the tachyon field by using the
latest data from observations. The conclusions are drawn in Sec.
III.

\section{Interacting holographic tachyon dark energy model}

The fact that the tachyon can act as a source of dark energy with
different potential forms have been widely discussed in the
literature \cite{bagla, ying, liddle, copeland4}. The tachyon can
be described by an effective field theory corresponding to a
tachyon condensate in a certain class of string theories with the
following effective action \cite{roo, sen}
\begin{equation}\label{eq21}
S=\int d^4x\sqrt{-g}\left[\frac{R}{16\pi
G}-V(\phi)\sqrt{1+g^{\mu\nu}\partial_{\mu}\phi\partial_{\nu}\phi}\right],
\end{equation}
where $V(\phi)$ is the tachyon potential and $R$ the Ricci scalar.
The physics of tachyon condensation is described by the above
action for all values of $\phi$ provided the string coupling and
the second derivative of $\phi$ are small. The corresponding
energy-momentum tensor of the tachyon field has the form
\begin{equation}\label{eq22}
T_{\mu\nu}=\frac{V(\phi)\partial_{\mu}\phi\partial_{\nu}\phi}{\sqrt{1+g^{\alpha\beta}\partial_{\alpha}\phi\partial_{\beta}\phi}}-g_{\mu\nu}V(\phi)\sqrt{1+g^{\alpha\beta}\partial_{\alpha}\phi\partial_{\beta}\phi}.
\end{equation}
In the flat FRW background the energy density $\rho_{t}$ and the
pressure $p_{t}$ are given by
\begin{equation}\label{eq23}
\rho_t=-T_{0}{}^{0}=\frac{V(\phi)}{\sqrt{1-\dot{\phi}^2}},
\end{equation}
\begin{equation}\label{eq24}
p_t=T_{i}{}^{i}=-V(\phi)\sqrt{1-\dot{\phi}^2},
\end{equation}
where no summation over repeated indices is assumed and the dot
stands for the derivative with respect to cosmic time.

From Eqs. (\ref{eq23}) and (\ref{eq24}) we obtain the tachyon
equation of state parameter
\begin{equation}\label{eq25}
w_t=\frac{p_{t}}{\rho_{t}}=\dot{\phi}^2-1.
\end{equation} In order to have a real energy density for the tachyon we require
$0<\dot{\phi^{2}}<1$ which implies, from Eq. (\ref{eq25}), that the
equation of state parameter is constrained to $-1<w_{t}<0$. Hence,
irrespective of the form of the potential, the tachyonic scalar
field cannot achieve an equation of state parameter that enters
the phantom regime.

In order to impose the holographic nature to the tachyon, we
should identify $\rho_{\rm t}$ with $\rho_{\Lambda}$. We consider
a spatially flat FRW universe filled with DM and HDE. The
Friedmann equation reads
\begin{equation}\label{Friedmanneq}
3M_{P}^2H^2=\rho_{\rm m}+\rho_{\rm t}.
\end{equation}
Given that the matter component is mainly contributed by the cold
dark matter and that it is generally assumed that baryons do not
interact with the dark sector, we shall ignore the contribution of
the baryon matter here for simplicity.

In the case of an interaction between HDE and DM, their energy
densities no longer satisfy independent conservation laws. They
obey instead
\begin{equation}
\dot{\rho}_m+3H\rho_m = Q \;,
\end{equation}
\begin{equation}
 \dot{\rho}_t+3H(1+\omega_t)\rho_t = -Q,
 \label{diff2}
\end{equation}where $Q$ is an interaction term whose form is not unique. Here in this letter
we consider the following form
\begin{equation}
Q=3b^{2}H(\rho_{\rm m}+\rho_{\rm t}),
\end{equation} where $b^{2}$ is the coupling constant and $3H$ is attached for dimensional consistency. This particular interaction term was first
introduced on phenomenological grounds in the study of a suitable
coupling between a quintessence scalar field and a pressureless
cold dark matter component in order to alleviate the coincidence
problem \cite{Steinhardt}. For a rationale of this particular form
of the interaction term see \cite{Wang:2007ak}.

The term $b^{2}$ gauges the intensity of the coupling, being
$b^{2}=0$ the absence of interaction. Apart from this, it measures
to what extent the different evolution of the DM due to its
interaction with the DE gives rise to a different expansion
history of the Universe. A positive $b^{2}$ corresponds to a decay
of DE into DM. In fact, it can be seen that the coincidence
problem is substantially alleviated in the IHDE model, unlike the
$\Lambda$CDM one which does not have this advantage
\cite{Interaction in perturbation dynamics}. Furthermore, its
observational signatures were recently investigated and this model
was found to be mildly favored over the $\Lambda$CDM one
\cite{Ma:2007pd}.

Combining the definition of HDE (3) and that of the future event
horizon (4) we take the derivative with respect to $x=\ln{a}$ and
obtain
\begin{equation}
\label{rhop1}
 \rho'_t\equiv\frac{d\rho_t}{dx}=-6M^2_pH^2\Omega_t(1-\frac{\sqrt{\Omega_{t}}}{c}),
\end{equation}
where $\Omega_t=\rho_{t}/(3M^2_pH^2)$. Given that, from the
definition of the Hubble parameter, $\dot{\rho}_t\equiv
d\rho_t/dt=\rho'_tH$ and making use of the Friedmann equation
(\ref{Friedmanneq}), Eq. (\ref{diff2}) can be written as
\begin{equation}
\label{rhop2}
    \rho'_t+3(1+w_t)\rho_t=-9M^2_pb^2H^2.
\end{equation}

Combining the last two equations, we are led to the equation of
state parameter of this IHDE model,
\begin{equation}
\label{omegt}
w_t=-\frac{1}{3}-\frac{2}{3}\frac{\sqrt{\Omega_t}}{c}-\frac{b^2}{\Omega_t}.
\end{equation}
This is the equation we shall use throughout this letter. However,
other authors \cite{Kim:2005at} argued that
\begin{equation}
\label{wdeff} w_t^{\mathrm{eff}}=w_t+\frac{b^2}{\Omega_t}
=-\frac{1}{3}-\frac{2}{3}\frac{\sqrt{\Omega_t}}{c}
\end{equation}
should be used instead but this issue is not settled yet.

We must mention, however, that when the interaction between dark
components is present, the situation may become somewhat ambiguous
because the equation of state parameter $w_{t}$ loses its ability
to classify dark energies definitely, owing to the fact that now
DE and DM are entangled. Under these conditions, concepts such as
quintessence or phantom are not as clear as usual. Even though, we
can still use these conceptions in an undemanding sense as the
interacting term is very weak according to observations.

Inserting Eq. (\ref{omegt}) into Eq. (\ref{rhop2}) and using the
definition of $\Omega_t$, we arrive at
\begin{equation}\label{hp1}
    \frac{H'}{H}=-\frac{\Omega'_t}{2\Omega_t}+\frac{\sqrt{\Omega_t}}{c}-1.
\end{equation}
On the other hand, replacing $\dot{H}=H'H$ and $p_t=w_t\rho_t$
into the derivative of the Friedmann equation with respect to
cosmic time $\dot{H}=-\frac{1}{2M^2_p}(\rho+p)$ (where $\rho$ and
$p$ are the total energy density and pressure respectively), we
have
\begin{equation}\label{hp2}
    \frac{H'}{H}=\frac{1}{2}\Omega_t+\frac{\Omega^{3/2}_t}{c}+\frac{3}{2}b^2-\frac{3}{2}.
\end{equation}
If we combine now last two equations, we find the evolution
equation for $\Omega_t$
\begin{equation}\label{omdp}
\frac{d \Omega_t}{dx}=
  \Omega_t(1-\Omega_t)\left(1+\frac{2\sqrt{\Omega_t}}{c}-\frac{3b^2}{1-\Omega_t}\right),
\end{equation}
which governs the whole dynamics of the IHDE model.

\begin{figure}
\psfrag{phi}{$\phi$}
\psfrag{b2}{$b^2$}
\psfrag{c}{$c$}
\psfrag{z}{$z$}
\includegraphics[scale=0.885]{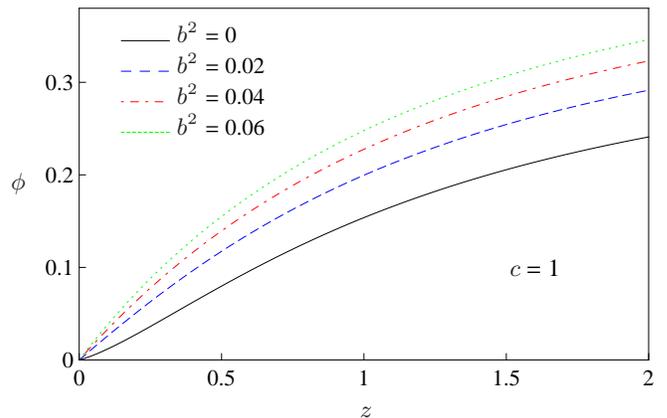}
\caption{The evolution of $\phi(z)$, where $\phi$ is in units of
$H_{0}^{-1}$, for a fixed $c$ and different values of the
coupling with $\Omega_{m0}=0.27$.}

\end{figure}

\begin{figure}
\psfrag{phi}{$\phi$} \psfrag{b2}{$b^2$} \psfrag{c}{$c$}
\psfrag{V}{$V$}
\includegraphics[scale=0.89]{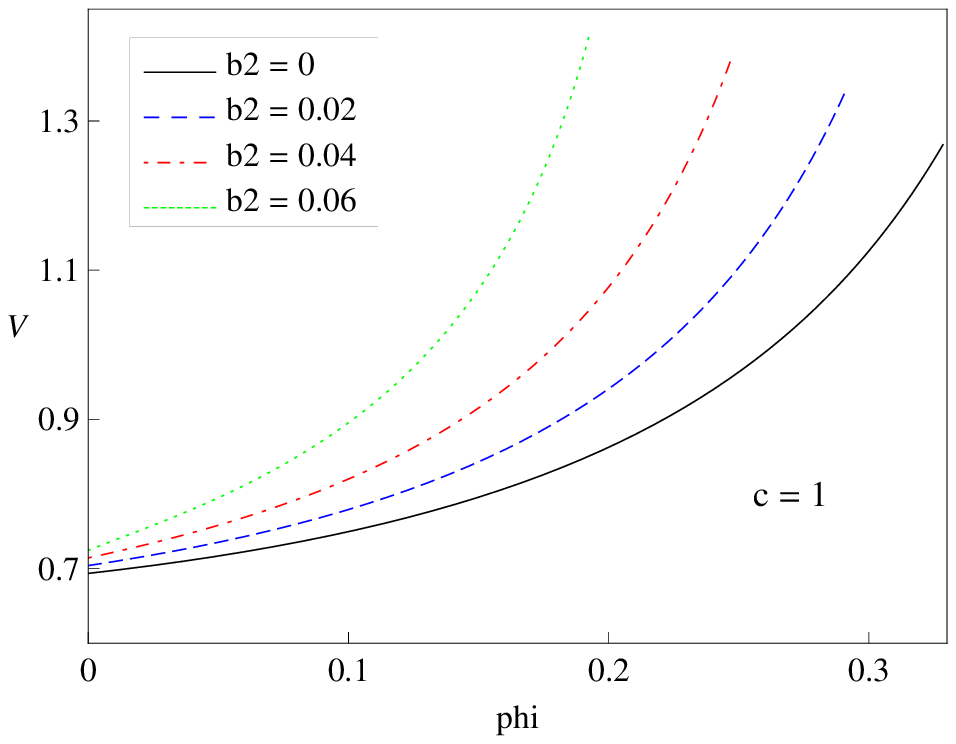}
\caption{The potential for the interacting holographic tachyon
model, where $\phi$ is in units of $H_{0}^{-1}$ and $V(\phi)$ in
$\varrho_{cr,0}$, for a fixed $c$ and different values of the
coupling. Here we have chosen $\Omega_{m0}=0.27$.}
\end{figure}

\begin{figure}
\psfrag{phi}{$\phi$} \psfrag{b2}{$b^2$} \psfrag{c}{$c$}
\psfrag{z}{$z$}
\includegraphics[scale=0.89]{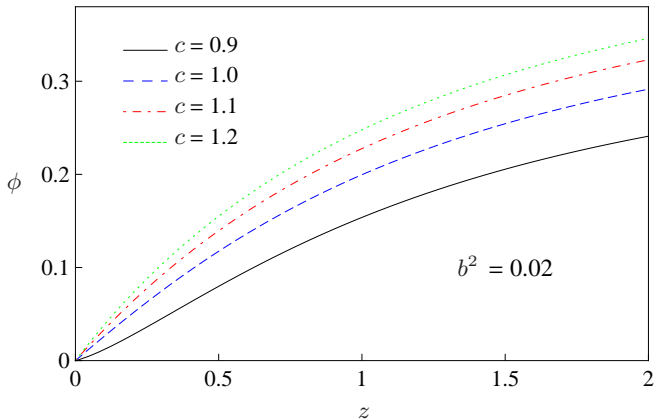}
\caption{The evolution of $\phi(z)$, where $\phi$ is in units of
$H_{0}^{-1}$, for a fixed coupling and different values of $c$.
As is usual, here we have considered $\Omega_{m,0}=0.27$.}
\end{figure}

\begin{figure}
\psfrag{phi}{$\phi$} \psfrag{b2}{$b^2$} \psfrag{c}{$c$}
\psfrag{V}{$V$}
\includegraphics[scale=0.89]{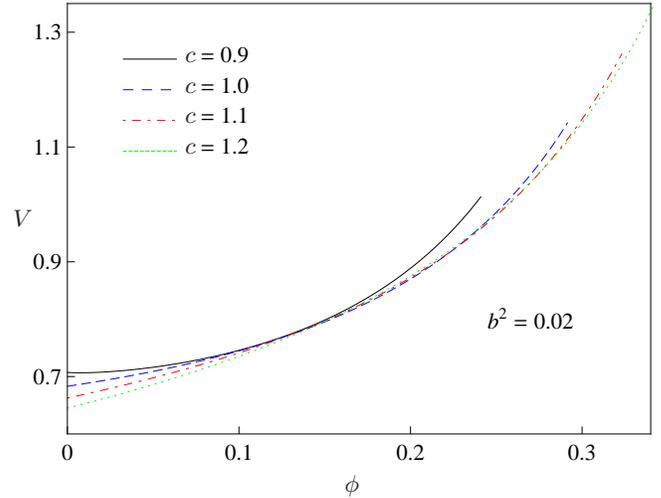}
\caption{The potential for the interacting holographic tachyon
model, where $\phi$ is in units of $H_{0}^{-1}$ and $V(\phi)$ in
$\varrho_{cr,0}$, for a fixed coupling and different values of
$c$ with $\Omega_{m,0}=0.27$.}
\end{figure}

\begin{figure}
\psfrag{z}{$z$}
\psfrag{q}{$q$}
\psfrag{b2}{$b^2$}
\includegraphics[scale=0.92]{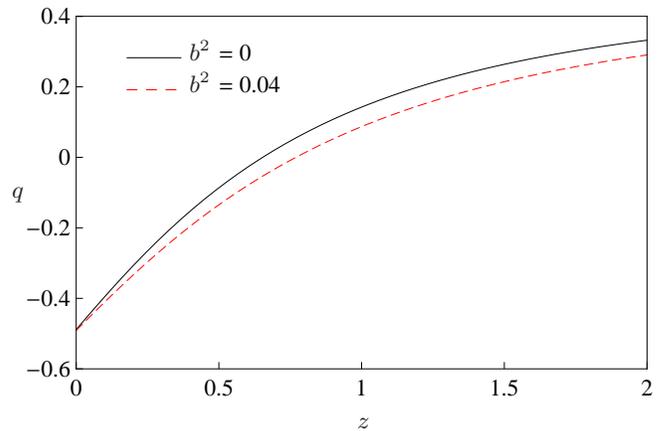}
\caption{Evolution of the deceleration parameter $q$ with and
without interaction for a fixed parameter $c=1$. We take here
$\Omega_{t,0}=0.73$.}
\end{figure}

\begin{figure}
\psfrag{Ot}{$\Omega_{t}$}
\psfrag{Om}{$\Omega_{m}$}
\psfrag{b2}{$\,b^2$}
\psfrag{z}{$z$}
\includegraphics[scale=0.96]{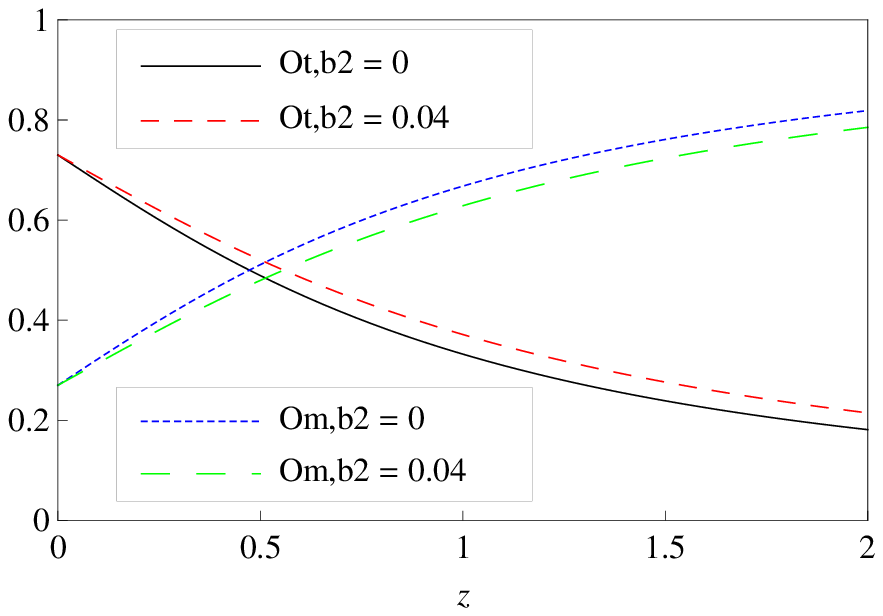}
\caption{Variation of $\Omega_{t}$ and $\Omega_{m}$ with respect
to the redshift for the holographic tachyon model with and without
interaction. We take in this plot $c=1$ and $\Omega_{t,0}=0.73$.}
\end{figure}

\begin{figure}
\psfrag{rhomb2}{$\rho_{m}$, $b^2$}
\psfrag{rhotb2}{$\rho_{t}$, $b^2$}
\psfrag{z}{$z$}
\includegraphics[scale=1]{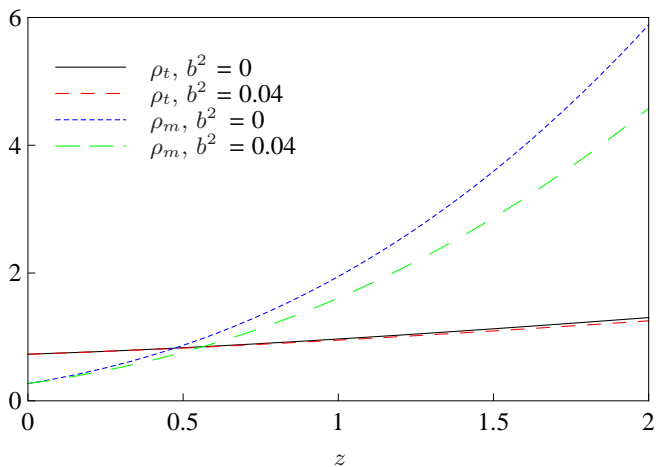}
\caption{Variation of $\rho_{t}$ and $\rho_{m}$ with respect to
$z$ in units of $\varrho_{cr,0}$ for the holographic tachyon model
with and without interaction. We take in this plot $c=1$ and
$\Omega_{t,0}=0.73$.}
\end{figure}
\begin{figure}
\psfrag{b2}{$b^2$} \psfrag{z}{$z$} \psfrag{r}{$r$}
\includegraphics[scale=0.96]{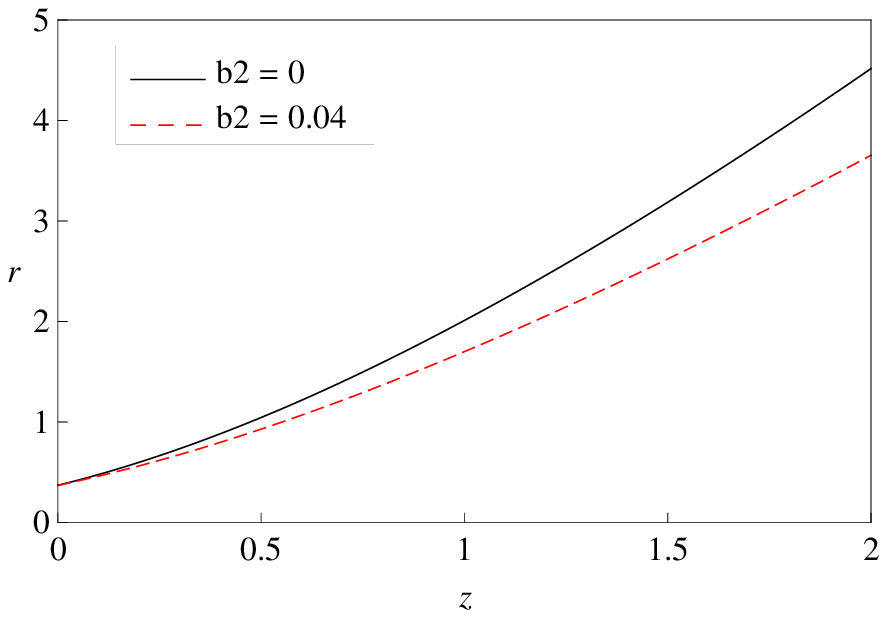}
\caption{Variation of the ratio $r\equiv\rho_m/\rho_t$
with respect to the redshift for the
holographic tachyon model with and without interaction. We take in
this plot $c=1$ and $\Omega_{t,0}=0.73$.}
\end{figure}

Since $\frac{d}{dt}=H\frac{d}{dx}=-H(1+z)\frac{d}{dz}$ we can
rewrite the above equation with respect to $z$ as
\begin{equation}\label{omdpz}
\frac{d \Omega_t}{dz}=-(1+z)^{-1}
  \Omega_t(1-\Omega_t)\left(1+\frac{2\sqrt{\Omega_t}}{c}-\frac{3b^2}{1-\Omega_t}\right).
\end{equation} Therefore, the differential equation for the Hubble
parameter $H(z)$ can be expressed as
\begin{equation}\label{Hz}
\frac{dH}{dz}=-(1+z)^{-1}
  H\left(\frac{1}{2}\Omega_t+\frac{\Omega^{3/2}_t}{c}+\frac{3}{2}b^2-\frac{3}{2}\right).
\end{equation}
The above equations can be solved numerically to obtain the
evolution of $\Omega_t$ and $H$ as a function of the redshift.

Using Eqs. (\ref{eq23}), (\ref{eq25}) and (\ref{Hz}), we derive
the interacting holographic tachyon potential

\begin{equation}\label{Vdephi}
\frac{V(\phi)}{\rho_{cr,0}}=H^{2}\Omega_{t}\sqrt{-w_{t}},
\end{equation}
where $\Omega_{t}$ and $w_{t}$ are respectively given by Eqs.
(\ref{omdpz}) and (\ref{omegt}), being $\rho_{cr,0}=3M^2_pH^2_0$
the critical energy density of the universe at the present epoch.
Besides, using Eqs. (\ref{eq25}) and (\ref{Hz}), the derivative of
the interacting holographic tachyon scalar field $\phi$ with
respect to the redshift $z$ can be expressed as

\begin{equation}\label{phiprime}
\frac{\phi'}{H_{0}^{-1}}=\pm\frac{\sqrt{1+w_{t}}}{H(1+z)}.
\end{equation} The sign is in fact arbitrary as it can be changed by a
redefinition of the field $\phi\rightarrow-\phi$.

The above equation cannot be solved analytically, however, the
evolutionary form of the interacting holographic tachyon field can
be easily obtained integrating it numerically from $z=0$ to a
given value $z$.

The field amplitude at the present epoch ($z=0$) is taken to
vanish, $\phi(0)=0$. Changing this initial value is equivalent to
a displacement in $\phi$ by a constant value $\phi_{0}=\phi(z=0)$,
which does not affect the shape of the field.

We note that Eqs. (\ref{Vdephi}) and (\ref{phiprime}) are formally
the same as in \cite{Zhang:2007es}, but $H(z)$ is different in our
case due to the interaction which modifies the expansion history
of the Universe.

 As already discussed in \cite{Wang:2004jf} the interaction
$Q$ is very weak and positive and the parameters $b^2$ and $c$ are
not totally free; they need to satisfy some constraints. Following
the latest observational results for the IHDE models
\cite{Feng:2007wn, Wu:2007fs, Ma:2007pd}, we take $0\le b^2\le
0.06$ and $\sqrt{\Omega_t}<c<1.255$, where the lower bound of $c$
comes from the second law of thermodynamics. The interaction
coupling has an upper limit because of the evolutionary behavior
of the HDE \cite{Interaction in the CMB multipoles}. As it can be
seen in Fig. 5, where the dependence of the deceleration parameter
$q = - \ddot{a}/(aH^2)$ on the coupling for a fixed $c$ is shown,
the interaction has an appreciable effect on the acceleration
history of the Universe. For a fixed parameter $c$, the cosmic
acceleration starts earlier for the cases with interaction than
the one without coupling as DE dominates earlier. This result was
also previously obtained by other authors \cite{Interacting HDE,
Interaction in perturbation dynamics, Interaction in the CMB
multipoles, Das:2005yj, Phenomenological interaction}. Moreover,
for larger coupling between DE and DM, the acceleration starts
earlier. However, the cases with smaller coupling will get larger
acceleration finally in the far future. Besides, the cases with a
fixed small $b^{2}$ and various values of $c$ are also
interesting. The Universe starts to accelerate earlier when $c$ is
larger for the same coupling $b^{2}$, but finally a smaller $c$
will lead to a larger acceleration \cite{Zhang:2007uh}.

The analytical form of the potential in terms of the interacting
holographic tachyon field cannot be determined due to the
complexity of the equations involved. Although, we can obtain it
numerically. The reconstructed $V(\phi)$ is plotted in Figs. 2
and 4. The scalar field $\phi(z)$ is also reconstructed by
solving Eq. (\ref{phiprime})
and shown in Figs. 1 and 3. Selected curves are plotted for the
cases of $c=1$ and $b^{2}= 0, 0.02, 0.04$ and $0.06$ in Figs. 1
and 2. And for the cases of $b^{2}=0.02$ and $c=0.9, 1.0, 1.1$ and
$1.2$ in Figs. 3 and 4. The present fractional matter density is
chosen to be $\Omega_{m,0}=0.27$. Figs. 1 to 4 display the
dynamics of the interacting tachyon scalar field explicitly.
Following the interacting holographic evolution of the Universe,
all the potentials are more steep in the early epoch, tending to
be flat near today. Consequently, the tachyon field $\phi$ rolls
down the potential more slowly as the Universe expands (the
kinetic term $\dot{\phi}^{2}$ gradually decreases) and the
equation of state parameter tends to negative values close to $-1$
according to Eq. (\ref{eq25}) as $\dot{\phi}\rightarrow0$. As a
result $dw_{t}/d$lna$<0$. Note that $\phi(z)$ increases with $z$
but becomes finite at high redshift. This means that $\phi$
decreases as the Universe expands.

Similar behavior has been obtained  in \cite{Zhang:2007es} for a
holographic tachyon model. This was to be expected because the
coupling that gauges the interaction in the IHDE model is small,
otherwise this model would deviate significantly from the
concordance model, making it incompatible with observations
\cite{matter perturbations}.

Fig. 6 shows the impact of the interaction between HDE and DM,
namely, $\Omega_t$ increases at a faster rate as compared to the
non-interacting case. In addition, from Fig. 7 we learn that the
point where $\rho_{t}$ and $\rho_{m}$ cross, $\rho_t=\rho_m$,
occurs earlier in the interacting scenario. This latter feature is
appreciated in more detail in Fig. 8 where the dependence of the
ratio $r \equiv \rho_m/\rho_t$ with respect to the redshift $z$ is
depicted.  The aforementioned ratio decreases monotonously with
the expansion and varies slowly at the present epoch, decreasing
slower when the interaction is considered. This implies that in
this scenario the coincidence problem gets alleviated and besides,
that DE is decaying into DM in recent epochs. Furthermore, the
different evolution of DM due to its interaction with DE also
gives rise to a different expansion history of the Universe.
Moreover, the standard structure formation scenario, with
$\rho_{m} \propto a^{-3}$, is altered when DM interacts with DE
due to a different evolution of the matter density perturbations.
These matter perturbations were studied in \cite{matter
perturbations, Interaction in perturbation dynamics}.

\section{Conclusions}

We have proposed an interacting holographic tachyon model of dark
energy with the future event horizon as infrared cut-off. This has
been done by establishing a correspondence between the IHDE model
and the tachyon field. We have also carried out a throughout
analysis of its evolution and deduce its cosmological
consequences.

By assuming that the scalar field models of dark energy are
effective theories of an underlying theory of dark energy and
regarding the scalar field model as an effective description of
such a theory, we can use the tachyon scalar field model to mimic
the evolving behavior of the IHDE. As a result, we have
reconstructed the interacting holographic tachyon model in the
region $-1<w<0$, i.e. before the phantom crossing, which is the
allowed region for the tachyon field.

In summary, we have shown that the interacting holographic
evolution of the universe can be described completely by a tachyon
scalar field and that the obtained results enter inside the valid
region of different experimental data. We must finally add that a
paper that deals with the interacting tachyon in the holographic
context appeared recently \cite{Micheletti:2009jy} but the
motivation and objectives in it are different from ours.

\begin{acknowledgments}
We are grateful to Prof. Pedro F. Gonz{\'a}lez-D{\'i}az for
reading the manuscript. This work was supported by DGICYT under
Research Project No.~FIS2005-01180, by the Spanish MICINN Project
FIS2008-06078-C03-03, by the Deutsche Forschungsgemeinschaft (DFG)
through SFB/TR7 ``Gravitationswellenastronomie'' and by DICYT
040831 CM, Universidad de Santiago de Chile.
\end{acknowledgments}

\end{document}